\documentclass[graybox]{svmult}

\usepackage{mathptmx}       
\usepackage{helvet}         
\usepackage{courier}        
\usepackage{type1cm}        
\usepackage{makeidx}         
\usepackage{graphicx}        
\usepackage{multicol}        
\usepackage[bottom]{footmisc}

\usepackage{amsmath}
\usepackage{amssymb}
\usepackage{undertilde}
\usepackage{wrapfig}
\renewcommand{\vec}{\overrightarrow}

\makeindex             

\begin{document}

\title*{Modelling of an Ionic Electroactive Polymer by the Thermodynamics of Linear Irreversible Processes}
\titlerunning{Modelling of an ionic electroactive polymer}
\author{M. Tixier and J. Pouget}
\institute{Mireille Tixier \at D\'{e}partement de Physique, Universit\'{e} de Versailles Saint Quentin,
45, avenue des Etats-Unis, F-78035 Versailles, France, \email{mireille.tixier@uvsq.fr}
\and Jo\"{e}l Pouget \at Sorbonne Universit\'{e}, UPMC Univ. Paris 6, UMR 7190, Institut Jean le Rond d'Alembert, F-75005 Paris, France; CNRS, UMR 7190, Institut Jean le Rond d'Alembert, F-75005 Paris, France, \email{joel.pouget@upmc.fr}}

\maketitle

\abstract*{Ionic polymer-metal composites consist in a thin film of electro-active polymers (Nafion$^{\textregistered}$ for example) sandwiched between two metallic electrodes. They 
can be used as sensors or actuators. The polymer is saturated with water, which 
causes a complete dissociation and the release of small cations. The strip 
undergoes large bending motions when it is submitted to an orthogonal electric field and vice 
versa. We used a continuous medium approach and a coarse grain model; the system is depicted
as a deformable porous medium in which flows an ionic solution. We write microscale balance 
laws and thermodynamic relations for each phase, then for the complete material using an average 
technique. Entropy production, then constitutive equations are deduced : a Kelvin-Voigt 
stress-strain relation, generalized Fourier's and Darcy's laws and a Nernst-Planck equation. We 
applied this model to a cantilever E.A.P. strip undergoing a continuous potential difference 
(static case); a shear force may be applied to the free end to prevent its displacement. 
Applied forces and deflection are calculated using a beam model in large displacements. 
The results obtained are in good agreement with the experimental data published in the literature.}


\abstract{Ionic polymer-metal composites consist in a thin film of electro-active polymers (Nafion$^{\textregistered}$ for example) sandwiched between two metallic electrodes. They 
can be used as sensors or actuators. The polymer is saturated with water, which 
causes a complete dissociation and the release of small cations. The strip 
undergoes large bending motions when it is submitted to an orthogonal electric field and vice 
versa. We used a continuous medium approach and a coarse grain model; the system is depicted
as a deformable porous medium in which flows an ionic solution. We write microscale balance 
laws and thermodynamic relations for each phase, then for the complete material using an average 
technique. Entropy production, then constitutive equations are deduced : a Kelvin-Voigt 
stress-strain relation, generalized Fourier's and Darcy's laws and a Nernst-Planck equation. We 
applied this model to a cantilever E.A.P. strip undergoing a continuous potential difference 
(static case); a shear force may be applied to the free end to prevent its displacement. 
Applied forces and deflection are calculated using a beam model in large displacements. 
The results obtained are in good agreement with the experimental data published in the literature.}

\section{Introduction}
\label{sec:1}

Electro-active polymers (EAP) have attracted much attention from scientists and engineers because 
of their very promising applications in many areas of science and the growing market. Their behavior and 
electro-chemical-mechanical interactions are of great interest and curiosity for research. In particular, the properties 
of these materials are highly attractive for biomimetic applications  (for instance, in robotic mechanisms are 
based on biologically inspired models), for the rise of artificial muscles (Bar-Cohen, 2001), and for haptic actuators. More recently 
EAPs are excellent candidates for energy harvesting devices (Brafau-Penella et al., 2008). Promising applications of this smart material 
consisting of micro-electromechanical systems (MEMS) at the sub-micron scale are also investigated for accurate medical 
control (Yoon et al., 2007). 

\noindent 
The purpose of the present study is to construct step by step a micro-mechanical model which accounts for  
couplings between the ion transport, electric field and elastic deformation in order to deduce the constitutive equations 
for this material. Next, an application to the actuation of beam made of thin layer of EAP is presented. Roughly speaking, 
an electro-active polymer (EAP) is a polymer that exhibits a mechanical response, such as stretching, contracting, or 
bending for example, when subject to an electric field (only few volts are needed for actuation). Conversely, the EAP can 
produce energy in response to a mechanical loading. 

\noindent 
The terminology electro-active polymer has very wide meaning and can be applied to a large category of materials. 
The electro-active polymers are generally divided in two main classes : (i) the electronic EAPs, in which activation is caused 
by electro-active force between both electrodes which squeezes the polymer and (ii) the ionic EAPs, in which actuation 
is due to the displacement of ions inside the polymer. Both classes are divided into subfamilies according to the 
physical or chemical principles of activation. The electronic EAP family encompasses ferroelectric polymers, electrets 
(PolyVinyliDene Fluoride (PVDF) is an example), dielectric elastomers, electroactive papers, liquide crystal polymers and 
many others. The ionic EAP category comprises ionic gels, ionic composite (IPMC) (such as Nafion$^{\textregistered}$ or 
Flemion$^{\textregistered}$), ionic conductor polymers (the strong conductivity is due to oxychloreduction process), nanotube 
of carbone (the electrolyte is modified by additional charges which produce volume change), 
electrorheologic fluid (fluid with micro particules changing the rheological properties of fluid, viscosity for instance) 
among others. The reader can refer to Bar-Cohen (2001) for many more details. Each category possesses their own 
advantages and their drawbacks. \\

\noindent
The present work addresses investigation of electro-active polymers belonging to ionic class and more precisely to ionic-exchange 
polymer-metal composite (IPMC because of the metallic electrodes on the layer faces). The latter consists in an ionic polymer 
(Nafion$^{\textregistered}$, for instance) sandwiched with two electrodes onto the upper and lower surfaces of the polymer layer. Katchalsky, on 1949 
was one of the first investigators to report the ionic chemo-mechanical deformation of polyectrolytes such as polyacrylic 
acid or polyvinyl chloride systems. More recently a great interest has been devoted to EAPs due to the similarities with biological 
tissues in terms of achievable stress and they are often called artificial muscles. Moreover, the material has potential applications 
in the field of robotics, medical technology and so on. Investigation on EAP have been traced to Shahinpoor and co-workers 
(Shahinpoor, 1994; Shahinpoor et al., 1998) and some many other researchers. 

\noindent
Modeling of EAPs must include complicated  electro-chemical-mechanical couplings. Different kinds of approaches have been 
proposed. Newbury and Leo (2001,
2002) developed empirical and heuristic models to explain sensing and 
actuating properties of ionic polymer benders. Model based on electrostatic interactions produced by ion motion has been 
developed by Todokoru et al. (2000). A model including the effect of electric field, current, pressure gradient and water flux 
as state variables has been proposed by de Gennes et al. (2000) using the concept of irreversible thermodynamics. A more 
sophisticated model nonetheless closer to the realistic properties was developed by Nemat-Nasser and Li (2000) and Nemat-Nasser
(2002). The model is based on the micro-mechanics of ionic polymers including ion transport. Finite element 3D 
model has been studied by Vokoun et al. (2015) to solve the basic governing physical equations for EAPs proposed by Nemat-Nasser 
with given boundary conditions.  A model of electro-viscoelastic polymers as an extension of the nonlinear electro-elasticity theory has been 
discussed by Ask et al.
(2012) and finite element numerical simulations were also presented. \\

\noindent
The proposed model accounts for electro-mechanical and chemical-electric coupling of ion transport, electric field and elastic deformation 
to produce the response of the EAP. We first investigate the conservation laws of the different phases at the micro level. An averaging 
statistical process  applied to the different phase quantities and to the equations of the conservation laws at the micro-scale is 
used to deduce, in a representative elementary volume containing all the phases, the equations of the conservation laws of the 
polymer at the continuum level. We write down conservation laws (mass, momentum, energy) in the framework of non-equilibrium  
thermodynamics. The thermodynamics of linear irreversible process allows us to identify the fluxes and generalized forces (Tixier and Pouget, 2014) 
and the constitutive equations for the continuum model are consequently obtained (Tixier and
Pouget, 2016). A generalized Darcy's law and the balance 
for ion flux (a kind of Nernst-Plank equation) are deduced from the thermodynamics relations and Gibbs relation. Along with the former 
equations the stress-strain equation and that of the electric charge conservation are also considered (Tixier and
Pouget, 2016).  \\

\noindent
The paper is organized as follows, the next section is devoted to the description of the EAP giving the main properties and the 
way of modeling the material. The section reports also the method used for arriving at the continuum model. The Sect. 3 
concerns the conservation laws, especially the energy balance laws.  The fundamental thermodynamic relations as well as 
the entropy production are reported in Sect.4. Gibbs', Euler's and Gibbs-Duhem's relations are given in order to deduce the rate 
of entropy production. The constitutive equations are presented in Sect. 5, in particular the stress-strain relation, Nernst-Plank 
equation and generalized Darcy's law are written in terms of concentration, electric field and pressure. The Sect. 6 proposes 
a validation of the model by studying the bending actuation of EAP layer subject to a constant difference of electric 
potential applied to the upper and lower electrodes. Comparisons to experimental results available in the literature ascertain 
the present model. The most pertinent results are summarized in the conclusions.

\section{Description and Modelling of the Material}
\label{sec:2}

The system we study is an ionic polymer-metal composite (IPMC); it consists of a membrane of 
polyelectrolyte coated on both sides with thin metal layers acting as electrodes. The polymer 
is saturated with water, which results in a quasi-complete dissociation : anions remain
bound to the polymer backbone, whereas small cations are released in water 
(Chabé, 2008). When an electric field perpendicular to the electrodes is
applied, cations are attracted by the negative electrode and carry solvent away by osmosis. 
As a result, the polymer swells near the negative electrode and contracts on the opposite side, 
leading to the bending of the strip.

\begin{figure*} [h]
 \includegraphics[width=0.50\textwidth]{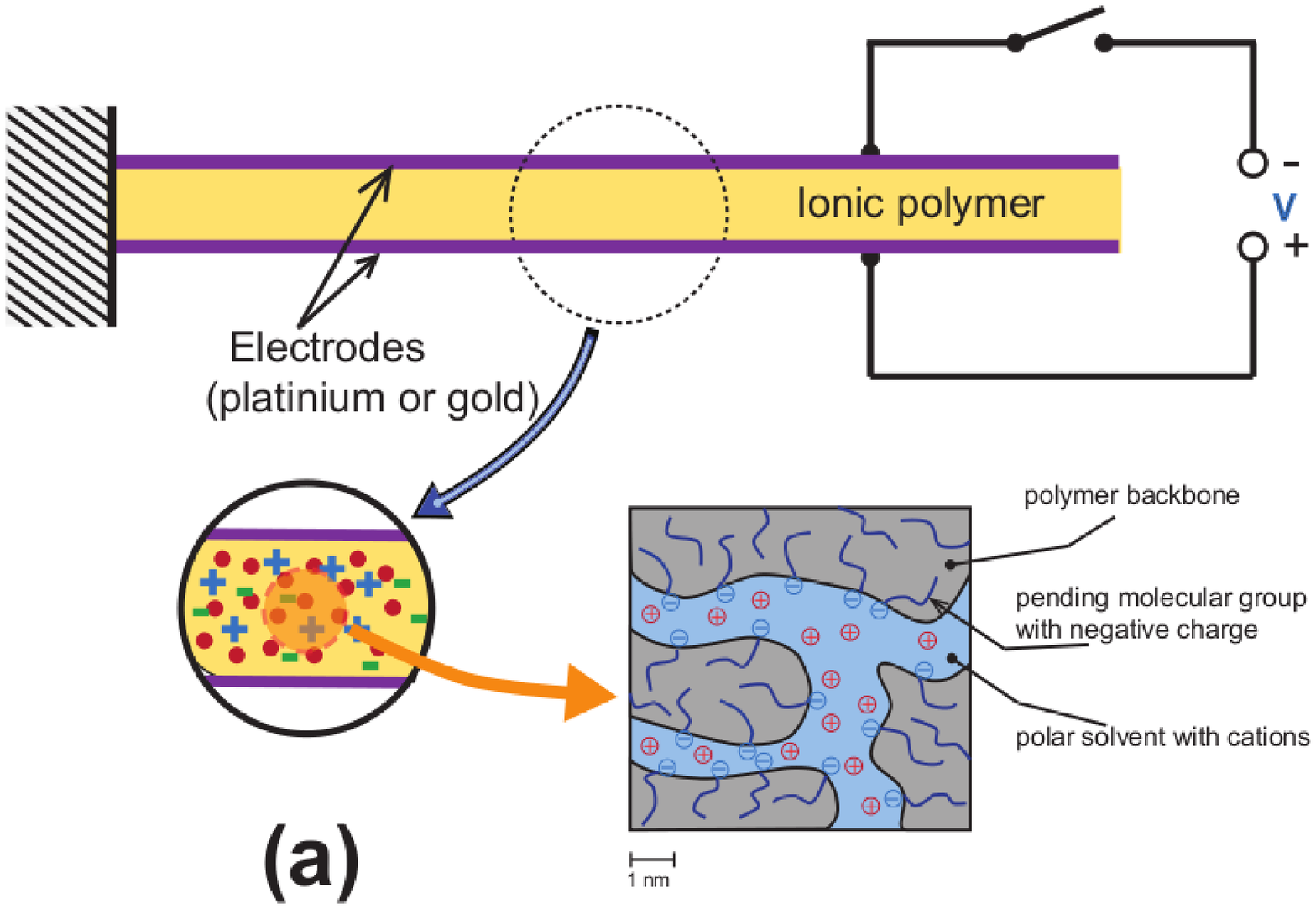}
 \includegraphics[width=0.49\textwidth]{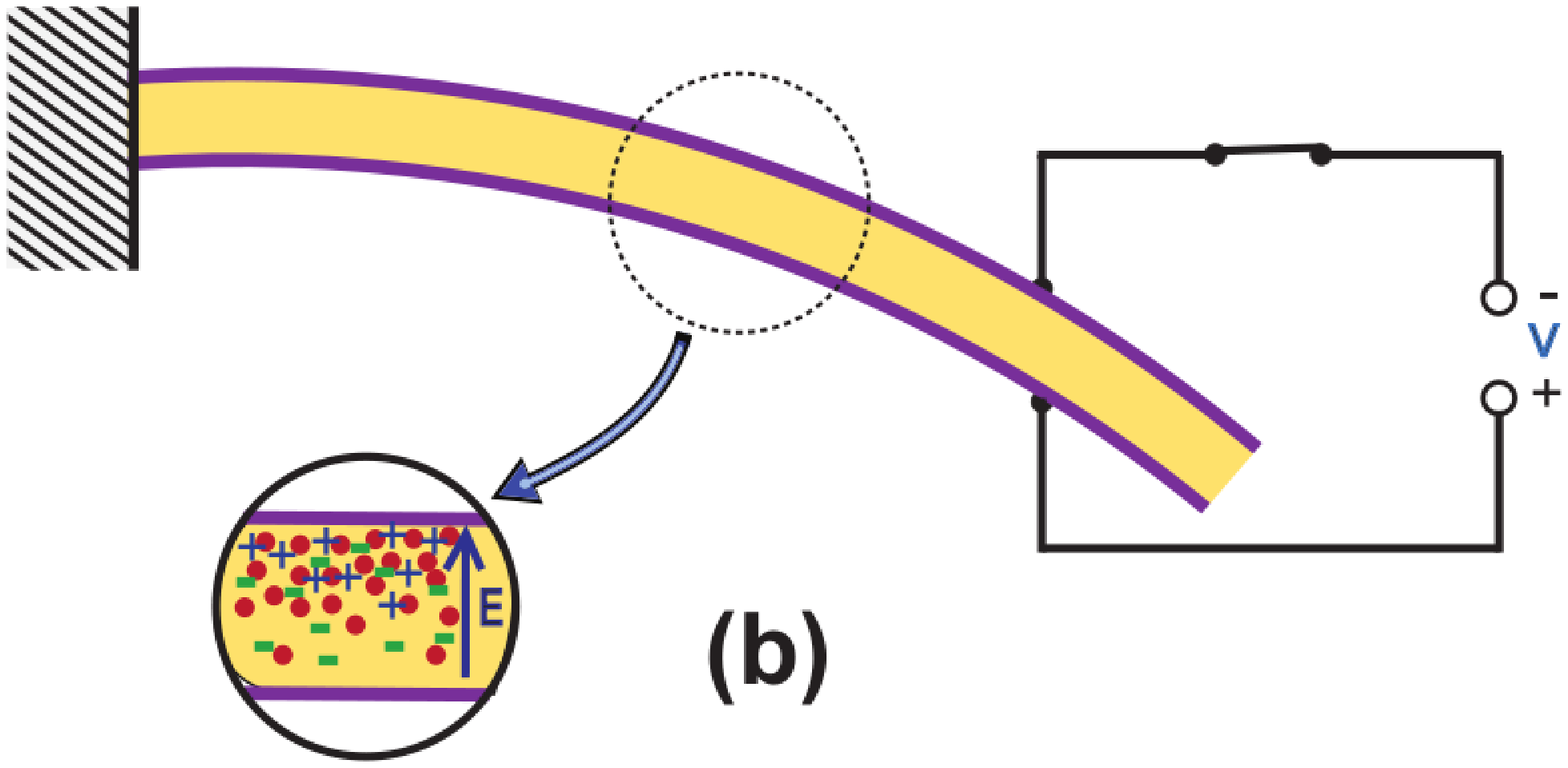}
\caption{Deformable porous medium : (a) Undeformed strip (b) Strip bending under an applied electric field }
\label{fig:1}
\end{figure*}

To model this system, the polymer chains are assimilated to a deformable porous medium
saturated by an ionic solution composed by water and cations. We suppose that the solution 
is dilute. We depicted the 
complete material as the superposition of three systems : a deformable solid made up of 
polymer backbone negatively charged, a solvent (the water) and cations (for schematic 
representation, see the insets a and b of Fig. 1). The solid and liquid phases
are assumed to be incompressible phases separated by an interface whose
thickness is supposed to be negligible. We identify the quantities relative
to the different components \ by subscripts : $1$ refers to cations, $2$ to
solvent, $3$ to solid, $i$ to the interface and $4$ to the solution, that is
both components $1$ and $2$; quantities without subscript refer to the whole material. 
The different components (except $1$) as well as the global material are assimilated to continua. 
We venture the hypothesis that gravity and magnetic field are negligible, so the only external 
force acting on the system is the electric force (Tixier and
Pouget, 2014).

\subsection{Average Process}
\label{subsec:21}

We describe this medium using a coarse-grained model developed for two-phase
mixtures (Drew, 1983; Drew and Passman, 1998; Ishii and Hibiki, 2006; Lhuillier, 2003; Nigmatulin, 1979, 1990).
We use two scales. The microscopic scale must be small enough so that the corresponding volume 
only contains a single phase ($3$ or $4$), but large enough to use a continuous medium model.  For 
Nafion$^{\textregistered}$ completely saturated with water, it is about hundred Angstroms. At the macroscopic scale, 
we define a representative elementary volume (R.E.V.) which contains the two phases; it must be 
small enough so that average quantities relative to the whole material can be 
considered as local, and large enough so that these averages are relevant.  Its characteristic length 
is about micron (Chabé, 2008; Collette, 2008; Gierke et al.,
1981).

A microscale Heaviside-like function of presence $\chi _{k}\left( 
\vec{r},t\right) $ has been defined for the phases $3$ and $4$%
\begin{equation}
\chi _{k}=1\;when\;phase\;k\;occupies\;point\;\vec{r}%
\;at\;time\;t,\quad \chi _{k}=0\;otherwise.
\end{equation}%
The function of presence of the interface is the Dirac-like function $\chi
_{i}=-\vec{\nabla}\chi _{k}\cdot \vec{n_{k}}$ (in $%
m^{-1} $) where $\vec{n_{k}}$ is the outward-pointing unit normal to
the interface in the phase $k$. $\left\langle {}\right\rangle _{k}$ denotes
the average over the phase $k$ of a quantity relative to the phase $k$ only.
The macroscale quantities relative to the whole material are obtained by
statistically averaging the microscale quantities over the R.E.V., that is
by repeating many times the same experiment. We suppose that this average,
denoted by $\left\langle {}\right\rangle $, is equivalent to a volume
average (ergodic hypothesis) and commutes with the space and time
derivatives (Leibniz’ and Gauss’ rules; Drew, 1983; Lhuillier, 2003). 
A macroscale quantity $g_{k}$ verifies%
\begin{equation}
g_{k}=\left\langle \chi _{k}g_{k}^{0}\right\rangle =\phi _{k}\left\langle
g_{k}^{0}\right\rangle _{k},
\end{equation}%
where $g_{k}^{0}$ is the corresponding microscale quantity and $\phi
_{k}=\left\langle \chi _{k}\right\rangle $ the volume fraction of the phase $%
k$. $g_{k}$ is relative to the total volume of the whole material. 
In the following, we use superscript $^{0}$ to indicate microscale
quantities; the macroscale quantities, which are averages defined all over
the material, are written without superscript.

\subsection{Interface Modelling}
\label{subsec:22}
In practice, contact area between phases $3$ and $4$ has a certain
thickness; extensive physical quantities vary from one bulk phase to the
other one. This complicated reality can be modelled by two uniform bulk
phases separated by a discontinuity surface $\Sigma $ whose localization is
arbitrary. Let $\Omega $ be a cylinder crossing $\Sigma $, whose bases are
parallel to $\Sigma $. We denote by $\Omega _{3}$ and $\Omega _{4}$ the
parts of $\Omega $ respectively included in phases $3$ and $4$.

The continuous quantities relative to the contact zone are identified by a
superscript $^{0}$ and no subscript. A microscale quantity per surface unit $%
g_{i}^{0}$ related to the interface is defined by

\begin{equation}
g_{i}^{0}=\lim\limits_{\Sigma \longrightarrow 0}\frac{1}{\Sigma }\left\{
\int_{\Omega }g^{0}dv-\int_{\Omega _{3}}g_{3}^{0}dv-\int_{\Omega
_{4}}g_{4}^{0}dv\right\},  \label{Def-i0}
\end{equation}%
where $\Omega _{3}$ and $\Omega _{4}$ are small enough so that $g_{3}^{0}$
and $g_{4}^{0}$ are constant. Its average over the R.E.V. is the volume
quantity $g_{i}$ defined by
\begin{equation}
g_{i}=\left\langle g_{i}^{0}\chi _{i}\right\rangle.
\end{equation}%
We arbitrarily fix the interface position in such a way that it has no mass density. 
The different phases do not interpenetrate, thus we can write on the interfaces
\begin{equation}
\vec{V_{1}^{0}}=\vec{V_{2}^{0}}=\vec{V_{3}^{0}} =\vec{V_{4}^{0}}=\vec{V_{i}^{0}},
\label{CMcl}
\end{equation}
where $\vec{V_{k}^{0}}$ denotes the local velocity of the phase $k$. Moreover, we 
neglect all the velocities fluctuations on the R.E.V. scale.

\subsection{Particule Derivatives and Material Derivative}
\label{subsec:23}
In order to write the balance equations, it is
necessary to calculate the variations of the extensive quantities following
the material motion. This raises a problem because the different phases
do not move with the same velocity : velocities of the solid and the solution
are a priori different. Let us consider an extensive quantity of density 
$g \left( \vec{r},t \right)$ relative to the whole material. We can define
particle derivatives following the motion of the solid $(\frac{d_{3}}{dt})$, 
the solution $(\frac{d_{4}}{dt})$ or the interface $(\frac{d_{i}}{dt})$
\begin{equation}
\frac{d_{k}g}{dt}=\frac{\partial g}{\partial t}+\boldsymbol{\nabla} g\cdot 
\vec{V_{k}}.
\end{equation}%

According to the theory developped by O. Coussy (Biot, 1977; Coussy, 1989,
1995), 
we are also able to define a derivative following the motion of the different phases 
of the medium. We will call it the "material derivative"
\begin{equation}
\rho \frac{D}{Dt}\left( \frac{g}{\rho }\right) =\sum\limits_{k=3,4,i}\rho
_{k}\frac{d_{k}} {dt} \left( \frac{g_{k}}{\rho _{k}}\right)%
=\sum\limits_{k=3,4,i}\frac{\partial g_{k}}{\partial t}+div\left( g_{k}%
\vec{V_{k}}\right),
\end{equation}%
where $\rho _{k}$ is the mass density of the phase $k$ and $g_{3}$, $g_{4}$ and 
$g_{i}$ the densities relative to the total actual volume attached to the solid, the 
solution and the interface, respectively
\begin{equation}
g=g_{3}+g_{4}+g_{i}.
\end{equation}%
This derivative must not be confused with the derivative $\frac{d}{dt}$ following the 
barycentric velocity $\vec{V}$
\begin{equation}
\rho \frac{D}{Dt}\left( \frac{g}{\rho }\right) =\rho \frac{d}{dt}\left( 
\frac{g}{\rho }\right)
-\sum\limits_{k}\left[ div\left( g_{k}\left( \vec{V}-\vec{V_{k}}\right) \right) \right].
\end{equation}

\subsection{Balance Laws}
\label{subsec:24}
The balance equation of an extensive microscale quantity $g_{k}^{0} 
\left( \vec{x},t\right) $ writes
\begin{equation}
\frac{\partial g_{k}^{0}}{\partial t}+div\left( g_{k}^{0}\vec
{V_{k}^{0}}\right) =-div\vec{A_{k}^{0}}+B_{k}^{0},
\end{equation}%
where $\vec{A_{k}^{0}}$ is the flux of $g_{k}^{0}$ linked to phenomena other than 
convection and $B_{k}^{0}$ the volume production of $g_{k}^{0}$ (source term). At the 
macroscopic scale
\begin{equation}
\frac{\partial g_{k}}{\partial t}+div\left( g_{k}\vec{V_{k}} \right) =-div\vec{A_{k}}+B_{k}
-\left\langle \vec{A_{k}^{0}}.\vec{n_{k}}\chi _{i}\right\rangle,
\end{equation}%
in which
\begin{equation}
\vec{A_{k}}=\left\langle \chi _{k}\vec{A_{k}^{0}}%
\right\rangle, \qquad \qquad \qquad B_{k}=\left\langle \chi
_{k}B_{k}^{0}\right\rangle.
\end{equation}

In the case of an interface, the balance law is according to Ishii and Hibiki (2006)
\begin{equation}
\frac{\partial g_{i}^{0}}{\partial t}+div_{s}\left( g_{i}^{0}\vec{V_{i}^{0}}\right) 
=\sum\limits_{3,4}\left[ g_{k}^{0}\left( \vec{V_{k}}-\vec{V_{i}^{0}}\right) .\vec{n_{k}}
+\vec{A_{k}^{0}}.\vec{n_{k}}\right] -div_{s}\vec{A_{i}^{0}}+B_{i}^{0},
\end{equation}%
where $\vec{A_{i}^{0}}$ is the surface flux of $g_{i}^{0}$ along the interface due to phenomena 
other than convection, $B_{i}^{0}$ the surface production of $g_{i}^{0}$ and $div_{s}$ 
the surface divergence operator. At the macroscopic scale
\begin{equation}
\frac{\partial g_{i}}{\partial t}+div\left( g_{i}\vec{V_{i}}\right) =\sum\limits_{3,4}
\left\langle \chi _{i}\vec{A_{k}^{0}}.\vec{n_{k}}\right\rangle -div\vec{A_{i}}+B_{i}.
\end{equation}
Note that $g_{i}$, $\vec{A_{i}}$ and $B_{i}$ are volume quantities.

For the complete material, we deduce by summation :
\begin{equation}
\rho \frac{D}{Dt}\left( \frac{g}{\rho }\right) =-div\vec{A}+B,
\end{equation}%
where
\begin{equation}
\vec{A}=\sum\limits_{3,4,i}\vec{A_{k}} \;, \qquad \qquad
\qquad B=\sum\limits_{3,4,i}B_{k}.
\end{equation}

\section{Conservation Laws}
\label{sec:3}
We now write the balance laws of the following quantities : mass, electric charge, 
linear momentum, the different energies (potential, kinetic, total and internal), 
entropy as well as Maxwell's equations. 
\subsection{Conservation of the Mass}
\label{subsec:31}
In the absence of chemical reaction, there is no volume production of mass and the 
only flux is due to convection. The macroscale mass continuity equation thus writes 
\begin{equation}
\frac{\partial \rho _{k}}{\partial t}+div\left( \rho _{k}\vec{V_{k}}\right) =0\qquad (k=2,3),
\qquad \qquad \frac{\partial \rho }{\partial t}+div\left( \rho \vec{V}\right)=0,
\end{equation}%
with
\begin{equation}
\rho _{1}=\phi _{4}CM_{1}, \qquad \qquad \rho _{k}=\phi _{k}\rho
_{k}^{0}\qquad (k=2,3), \qquad \qquad \rho _{4}=\rho _{2}+\phi _{4}CM_{1},
\end{equation}%
$M_{k}$ is the molar mass of the component $k$ and $C$ the cations molar concentration 
relative to the liquid phase. We assume the concentration fluctuations are negligible.
\subsection{Electric Equations}
\label{subsec:32}
As in the case of mass, there is, for the electric charge, neither source term nor 
flux except convection. The electric charge conservation law can be written

\begin{equation}
div\vec{I}+\frac{\partial \rho Z}{\partial t}=0,
\end{equation}%
where
\begin{equation}
\rho Z=\sum\limits_{3,4}\rho _{k}Z_{k}+Z_{i}, \qquad \qquad 
\vec{I}=\sum\limits_{3,4,i}\vec{I_{k}},
\end{equation}
and
\begin{equation}
\vec{I_{3}}=\rho _{3}Z_{3}\vec{V_{3}}, \qquad \qquad \vec{%
I_{4}}=\rho _{1}Z_{1}\vec{V_{1}}.
\end{equation}%
$Z_{k}$ denotes the electric charge per unit of mass and $\vec{%
I_{k}}$ the current density vector of phase or interface $k$.

The electric field $\vec{E_{k}^{0}}$ and the electric displacement $\vec{D_{k}^{0}}$ 
verify Maxwell's equations and their boundary conditions. We assume that the electric field 
fluctuations are negligible and that macroscale electric fields are identical in all the 
phases. We thus obtain Maxwell's equations for the complete material
\begin{equation}
\begin{tabular}{lll}
$\vec{rot}\vec{E}=\vec{0}, \qquad \qquad $ & $%
div\vec{D}=\rho Z \;, \qquad \qquad $ & $\vec{D}=\varepsilon \vec{E}$, %
\end{tabular} \label{Maxwell}
\end{equation} 
where the permittivity $\varepsilon$ of the whole material writes
\begin{equation}
\varepsilon =\sum\limits_{3,4}\phi _{k}\left\langle \varepsilon
_{k}^{0}\right\rangle _{k}.
\end{equation}%
We conclude that the E.A.P. behaves like an isotropic homogeneous linear dielectric.
However its permittivity varies over time and space because of variations of the 
volume fractions $\phi_{k}$.

\subsection{Linear Momentum Conservation Law}
\label{subsec:33}
Since the gravity and the magnetic field are negligible, the only force applied is 
the electric force (source term). The linear momentum flux is related to the stress tensors 
$\utilde{\sigma} _{k}^{0}$ of the different phases.
From (\ref{CMcl}), we deduce that the linear momentum of the interface 
is zero, which leads to the following linear momentum conservation law for the interfaces
\begin{equation} 
\sum\limits_{3,4}{\left\langle \utilde{\sigma
_{k}^{0}} \cdot\vec{n_{k}}\chi _{i}\right\rangle} = Z_{i}\vec{E_{i}}.
\end{equation}
We obtain for the complete material
\begin{equation}
\rho \frac{D\vec{V}}{Dt}=\vec{div}\utilde{\sigma }+\rho Z\vec{E}
=\vec{div}\left[ \utilde{\sigma }+\varepsilon \left( \vec{E} \otimes \vec{E}-
\frac{E^{2}}{2}\utilde{1}\right) \right] +\frac{E^{2}}{2}\vec{grad}\varepsilon,  \label{CQ}
\end{equation}
using Maxwell's equations (\ref{Maxwell}). $\utilde{1}$ denotes the identity tensor.
\begin{equation}
\utilde{\sigma}=\sum\limits_{k=3,4}\utilde{\sigma_{k}},
\end{equation}
is a symmetric tensor. $\varepsilon \left( \vec{E}\otimes \vec{E}-\frac{E^{2}}{2}\utilde{1}
\right)$ is the Maxwell's tensor, which is here symmetric. The additional term $\frac{E^{2}}
{2}\vec{grad}\varepsilon $ results from the non homogeneous material permittivity.

\subsection{Energy Balance Laws}
\label{subsec:34}

\subsubsection{Potential Energy Balance Equation}
Potential energy production is equal to the volume power $-\vec{E_{k}^{0}}\cdot\vec{I_{k}^{0}}$ 
of the force due to the action of the electric field on the electric charges. The two 
phases are supposed to be non-dissipative isotropic linear media. As a consequence the balance 
equation for the potential energy (Poynting's theorem) can be written in the integral form 
(Jackson, 1975; Maugin, 1988)
\begin{equation}
\frac{d}{dt}\int_{\Omega }\frac{1}{2}\left( \vec{E}\cdot\vec{D}+\vec{B}\cdot\vec{H}\right) dv
=-\oint\nolimits_{\partial \Omega }\left( \vec{E}\wedge \vec{H}\right) \cdot
\vec{n}ds-\int_{\Omega }\vec{E}\cdot\vec{I}dv.
\end{equation}%
The left hand side represents the variation of the potential energy attached to the volume 
$\Omega $ following the charge motion. If the charges are mobile, the associated local 
equation is written as follows for the phase $k$, neglecting the magnetic field
\begin{equation}
\frac{\partial E_{pk}^{0}}{\partial t}+div\left( E_{pk}^{0}\vec{%
V_{k}^{0}}\right) =-\vec{E_{k}^{0}}\cdot\vec{I_{k}^{0}}%
\qquad \qquad k=3,4,  \label{Epm}
\end{equation}%
in which $E_{pk}^{0}=\frac{1}{2}\vec{D_{k}^{0}}\cdot\vec{E_{k}^{0}}$ is the potential energy 
per unit of volume of the phase $k$. The potential energy balance equation for the whole 
material is then
\begin{equation}
\rho \frac{D}{Dt}\left( \frac{E_{p}}{\rho }\right) =-\vec{E}\cdot \vec{I}, \label{Ep}
\end{equation}
where $E_{p}=\frac{1}{2}\vec{D}\cdot\vec{E}$.

\subsubsection{Kinetic Energy Balance Equation}

The relative velocities of the different phases are negligible compared to the 
velocities measured in the laboratory reference frame. Let's take for example a
strip of Nafion$^{\textregistered}$ which is $\textit{200}\;\mu m$ thick and $\textit{1.57}\;cm$ long, 
bending in an electric field; the tip displacement is about $\textit{4}\;mm$ and is 
obtained in $\textit{1}\;s$ (Nemat-Nasser and Li,
2000). Relative velocities of the order 
of $\mathit{2}\;\mathit{10}^{-4}\;m\;s^{-1}$ and absolute velocities close to 
$\mathit{4}\;\mathit{10}^{-3}\;m\;s^{-1}$ are thus obtained (Tixier and Pouget, 2016). In a first 
approximation, we can identify the kinetic energy $E_{c}$ of the complete material and 
the sum of the kinetic energies of the constituents. The kinetic energy balance equation 
derives from the linear momentum balance equation
\begin{equation}
\rho \frac{D}{Dt}\left( \frac{E_{c}}{\rho }\right) =\sum\limits_{3,4}\left[
div\left( \utilde{\sigma_{k}}\cdot \vec{V_{k}}\right) -\utilde{\sigma_{k}}:
\utilde{grad}\vec{V_{k}}\right] +\left( \vec{I}-\vec{i}\right) \cdot \vec{E},
\end{equation}%
where the diffusion current of the cations in the solution and of the interfaces is
\begin{equation}
\vec{i}=\vec{I}-\sum\limits_{k=3,4}\left( \rho_{k}Z_{k}\vec{V_{k}}\right) -Z_{i}\vec{V_{i}}.
\end{equation}

\subsubsection{Total Energy Balance Equation}
The total energy $E$ is the sum of internal, potential and kinetic energies. The source term is 
null, and the flux comes from contact forces work and heat conduction $\vec{Q}$. The total energy 
conservation law for the whole material is
\begin{equation}
\rho \frac{D}{Dt}\left( \frac{E}{\rho }\right) =div\left( \sum\limits_{k=3,4}%
\utilde{\sigma_{k}}\cdot \vec{V_{k}}\right) -div\vec{Q}.
\end{equation}%

\subsubsection{Internal Energy Balance Equation}
Internal energy $U$ is the difference between total energy and potential and kinetic energies 
\begin{equation}
U=E-E_{c}-E_{p},
\end{equation}%
which leads to 
\begin{equation}
\rho \frac{D}{Dt}\left( \frac{U}{\rho }\right) =\sum\limits_{3,4}\left(\utilde{\sigma_{k}}:
\utilde{grad}\vec{V_{k}}\right) +\vec{i}\cdot \vec{E}-div\vec{Q}.
\end{equation}%

\subsubsection{Interpretation of the Equations}
The energy balance laws we have written are relative to a thermodynamic closed system 
because of the use of the material derivative. Source terms correspond to 
conversion of one kind of energy into another one. Energy exchanges are summarized 
in the table \ref{tab:1}. Fluxes can be considered as the rate of variation of the 
quantity associated with the 
conduction phenomenon. The kinetic energy flux is equal to the work of the contact forces, 
the internal energy flux is the heat flux and the total energy flux is the sum of the two. 
We verify that there is no source term in this last equation. $\left( \vec{I}-\vec{i}\right) 
\cdot \vec{E}$ results in a motion of the electric charges subject to the electric field and 
is a conversion of potential energy into kinetic energy. $\vec{i}\cdot \vec{E}$ can be seen 
as Joule heating, which is a conversion of potential energy into internal energy.
$\sum\limits_{3,4}\left( \utilde{\sigma_{k}}:\utilde{grad}\vec{V_{k}}\right) $ represents 
the viscous dissipation, which converts kinetic energy into heat.

\begin{table}
\caption{Energy exchanges}
\label{tab:1}
\begin{tabular}{ccccc}
\hline
& flux & $E_{c}\longleftrightarrow E_{p}$ & $U\longleftrightarrow E_{p}$ & $%
E_{c}\longleftrightarrow U$ \\ \hline
$E_{p}$ &  & $-\left( \vec{I}-\vec{i}\right) \cdot \vec{E}$ & $-\vec{i}\cdot \vec{E}$ &  \\ 
$E_{c}$ & $\sum\limits_{3,4}\utilde{\sigma_{k}}\cdot \vec{V_{k}} $ & $+\left( \vec{I}-%
\vec{i}\right) \cdot \vec{E}$ &  & $-\sum\limits_{3,4}
\utilde{\sigma_{k}}:\utilde{grad}\vec{V_{k}}$ \\ 
$U$ & $ -\vec{Q} $ &  & $+\vec{i}\cdot \vec{E}$ & 
$+\sum\limits_{3,4}\left( \utilde{\sigma_{k}}:\utilde{grad}\vec{V_{k}}\right) $ \\ 
$E$ & $ \sum\limits_{3,4}\utilde{\sigma_{k}}\cdot \vec{V_{k}}-\vec{Q} $ &  &  &  \\ \hline
\end{tabular}%
\end{table}%

\section{Entropy Production}
\label{sec:4}
We shall now write the thermodynamic relations of the electroactive polymer. The thermodynamics 
of linear irreversible processes will allow us to identify the fluxes and the generalized forces 
(Tixier and Pouget, 2016).

\subsection{Entropy Balance Law}
\label{subsec:41}
The entropy balance law of the whole material is written as
\begin{equation}
\rho \frac{D}{Dt}\left( \frac{S}{\rho }\right) =s-div\vec{\Sigma},
\end{equation}%
where $S$, $\vec{\Sigma}$ and $s$ denote the entropy, the entropy flux vector 
and the rate of entropy production, respectively.

\subsection{Fundamental Thermodynamic Relations}
\label{subsec:42}
According to de Groot and Mazur (1962), the Gibbs relations of a single constituent 
solid and of a two-constituent fluid can be written at the microscopic scale
\begin{equation}
\begin{tabular}{l}
$\rho _{3}^{0}\frac{d_{3}^{0}}{dt}\left( \frac{U_{3}^{0}}{\rho _{3}^{0}}%
\right) =p_{3}^{0}\frac{1}{\rho _{3}^{0}}\frac{d_{3}^{0}\rho _{3}^{0}}{dt}+%
\utilde{\sigma _{3}^{0e}}^{s}:\frac{d_{3}^{0}\utilde{\varepsilon _{3}^{0}}%
^{s}}{dt}+\rho _{3}^{0}T_{3}^{0}\frac{d_{3}^{0}}{dt}\left( \frac{S_{3}^{0}}{%
\rho _{3}^{0}}\right) $, \\ 
$\frac{d_{4}^{0}}{dt}\left( \frac{U_{4}^{0}}{\rho _{4}^{0}}\right) =T_{4}^{0}%
\frac{d_{4}^{0}}{dt}\left( \frac{S_{4}^{0}}{\rho _{4}^{0}}\right) -p_{4}^{0}%
\frac{d_{4}^{0}}{dt}\left( \frac{1}{\rho _{4}^{0}}\right)
+\sum\limits_{k=1,2}\mu _{k}^{0}\frac{d_{4}^{0}}{dt}\left( \frac{\rho
_{k}^{\prime }}{\rho _{4}^{0}}\right) $, %
\end{tabular}%
\end{equation}%
where $T_{k}^{0}$ denotes the absolute temperature, $\utilde{\epsilon _{3}^{0}}$ 
and $\utilde{\sigma _{3}^{0e}}$ the strain tensor and the equilibrium stress tensor, 
$\utilde{\epsilon _{3}^{0s}}$ and $\utilde{\sigma _{3}^{0es}}$
the strain and stress deviator tensors. The solid pressure $p_{3}^{0}$ is related 
to the stress tensor and verifies the Euler's relation, as well as the liquid pressure 
$p_{4}^{0}$
\begin{equation}
\begin{tabular}{l}
$p_{3}^{0}=-\frac{1}{3}tr\left( \utilde{\sigma_{3}^{0e}}\right)
=T_{3}^{0}S_{3}^{0}-U_{3}^{0}+\mu _{3}^{0}\rho _{3}^{0}$, \\ 
$U_{4}^{0}-T_{4}^{0}S_{4}^{0}+p_{4}^{0}=\mu _{1}^{0}CM_{1} +\mu _{2}^{0}\frac{\rho
_{2}^{0}\phi _{2}}{\phi _{4}} $, %
\end{tabular}%
\end{equation}
$\mu _{k}^{0}$ is the chemical potential per unit of mass. We venture the hypothesis 
that the fluctuations over the R.E.V. of the intensive thermodynamic 
quantities (pressures, temperatures and chemical potentials) and of the strain and 
equilibrium stress tensors are negligible. We also suppose that the solid deformations 
are small. Making the hypothesis of local thermodynamic equilibrium, we derive
\begin{equation}
\begin{tabular}{l}
$p=p_{3}=p_{4}=p_{3}^{0}=p_{4}^{0}$, \\ 
$T=T_{3}=T_{4}=T_{i}=T_{3}^{0}=T_{4}^{0}.$%
\end{tabular}%
\end{equation}%
We thus obtain Gibbs, Euler's and Gibbs-Duhem relations of the whole material
\begin{equation}
\begin{tabular}{l}
$T\frac{D}{Dt}\left( \frac{S}{\rho }\right) =\frac{D}{Dt}\left( \frac{U}{%
\rho }\right) +p\frac{D}{Dt}\left( \frac{1}{\rho }\right) -\frac{1}{\rho }%
\utilde{\sigma _{3}^{e}}^{s}:\frac{d_{3}\utilde{\epsilon _{3}}^{s}}{dt}%
-\sum\limits_{1,2}\mu _{k}\frac{\rho _{4}}{\rho }\frac{d_{4}}{dt}\left( 
\frac{\rho _{k}}{\rho _{4}}\right) $, \\ 
$p=TS-U+\sum\limits_{k=1,2,3}\mu _{k}\rho _{k}$  \;, \\ 
$\frac{dp}{dt}=S\frac{dT}{dt}+\sum\limits_{k=1,2,3}\rho _{k}\frac{d\mu _{k}}{%
dt}-\utilde{\sigma ^{e}}^{s}:\utilde{grad}\vec{V}$. %
\end{tabular}%
\end{equation}%

\subsection{Entropy Production}
\label{subsec:43}
The stress tensor is composed of the equilibrium stress tensor $\utilde{\sigma ^{e}}$ 
and the viscous stress tensor $\utilde{\sigma ^{v}}$; this second part vanishes at
equilibrium
\begin{equation}
\utilde{\sigma }=\utilde{\sigma ^{e}}+\utilde{\sigma ^{v}}
=-p\utilde{1}+\utilde{\sigma ^{e}}^{s}+\utilde{\sigma ^{v}}.
\end{equation}%
Combining the internal energy and entropy equations with the Gibbs relation, the rate 
of entropy production $s$ can be identified
\begin{equation}
s=\frac{1}{T}\utilde{\sigma ^{v}}:\utilde{grad}\vec{V}+\frac{1}{T}%
\vec{i^{\prime }}\cdot\vec{E}-\frac{1}{T^{2}}\vec{Q^{\prime }}\cdot%
\vec{grad} T+\sum\limits_{k=1,2,3}\rho _{k}\left( \vec{V}%
-\vec{V_{k}}\right) \cdot\vec{grad}\frac{\mu _{k}}{T},
\end{equation}
with
\begin{equation}
\begin{tabular}{l}
$\vec{i^{\prime }}=\vec{I}-\rho Z\vec{V}$, \\ 
$\vec{Q^{\prime }}=\vec{Q}-\sum\limits_{k=3,4}U_{k} \left( \vec{V}-\vec{V_{k}}\right)
-\sum\limits_{k=3,4} \utilde{\sigma _{k}}\cdot\left( \vec{V_{k}}- \vec{V}\right) $. %
\end{tabular}%
\end{equation}

\subsection{Generalized Forces and Fluxes}
\label{subsec:44}
We define the mass diffusion flux of the cations in the solution $\vec{J_{1}}$ and 
the mass diffusion flux of the solution in the solid $\vec{J_{4}}$
\begin{equation}
\vec{J_{1}}=\rho _{1}\left( \vec{V_{1}}-\vec{V_{2}}\right), \qquad \qquad 
\qquad \vec{J_{4}}=\rho _{4}\left( \vec{V_{4}}-\vec{V_{3}}\right).
\end{equation}%
These two fluxes are linearly independant. The diffusion current 
$\vec{i^{\prime }}$ and the fluxes $\rho _{k}\left( \vec{V_{k}}-\vec{V}\right)$ 
can be expressed as functions of these two fluxes. We thus identify the fluxes along 
with the associated generalized forces (table \ref{tab:2}).
\begin{table}
\caption{Generalized fluxes and forces}
\label{tab:2}
\begin{tabular}{|l|l|}
\hline
Fluxes & Forces \\ \hline
$\frac{1}{3}tr \utilde{\sigma ^{v}} $ & $\frac{1}{T}div\vec{V}$ \\ \hline
$\vec{Q^{\prime }}$ & $\vec{grad}\frac{1}{T}$ \\ \hline
$\vec{J_{1}}$ & $\frac{\rho _{2}}{\rho _{4}}\left[ \frac{1}{T} Z_{1}\vec{E}-
\vec{grad}\frac{\mu _{1}}{T}+\vec{grad}\frac{\mu _{2}}{T}\right] $ \\ \hline
$\vec{J_{4}}$ & $\frac{\rho _{3}}{\rho }\left[ \frac{1}{T}\left(\frac{\rho _{1}}
{\rho _{4}}Z_{1}-Z_{3}\right) \vec{E}-\frac{\rho_{1}}{\rho _{4}}\vec{grad}
\frac{\mu _{1}}{T}-\frac{\rho _{2}}{\rho _{4}}\vec{grad}\frac{\mu _{2}}{T}+
\vec{grad}\frac{\mu _{3}}{T}\right] $ \\ \hline
$\utilde{\sigma ^{v}}^{s}$ & $\frac{1}{T} \left[ \frac{1}{2}\left( \utilde{grad}
\vec{V}+\utilde{grad}\vec{V}^{T}\right) -\frac{1}{3} \left( div\vec{V} \right) 
\utilde{1} \right] $ \\ \hline
\end{tabular}
\end{table}

\section{Constitutive Equations}
\label{sec:5}
We venture the hypothesis that the medium is isotropic. According to Curie symmetry 
principle, there can not be any coupling between fluxes and forces whose tensorial 
ranks differs from one unit. Moreover, we suppose that coupling between fluxes and 
different tensorial rank forces are negligible, which is commonly admitted (de Groot and Mazur, 1962).
We thus obtain a tensorial law (the rheological equation) and three vectorial constitutive 
equations.

\subsection{Rheological Equation}
\label{subsec:51}
Considering the symmetry of the tensor $\utilde{\sigma ^{v}}$, the scalar and tensorial 
fluxes are linear functions of the corresponding forces. Assuming that 
the complete material satisfies the Hooke's law at equilibrium and the 
liquid phase is newtonian and stokesian, the pressure verify
\begin{equation}
p=-\frac{1}{3}tr\left( \utilde{\sigma ^{e}}\right)
=\left( \lambda +\frac{2}{3}G\right) tr\utilde{\epsilon },
\end{equation}%
where $\lambda $ and $G$ denote the first Lam\'{e} constant and the shear
modulus of the complete material, respectively, and $\utilde{\epsilon }$
the material strain
\begin{equation}
\utilde{\epsilon }=\frac{1}{2}\left( \utilde{grad}\vec{u}%
+\utilde{grad}\vec{u}^{T} \right) \qquad \text{and}\qquad %
\overset{\bullet }{\utilde{\epsilon }}=\frac{1}{2}\left( %
\utilde{grad}\vec{V}+\utilde{grad}\vec{V}^{T}\right).
\end{equation}%
$\vec{u}$ is the displacement vector. The stress tensor of the complete material 
thus identifies with a Kelvin-Voigt model
\begin{equation}
\utilde{\sigma }=\lambda \left( tr\utilde{\epsilon }\right) \utilde{1}+2G
\utilde{\epsilon }+\lambda_{v} tr\overset{\bullet }{\utilde{\epsilon }} 
\utilde{1}+\mu_{v} \overset{\bullet } {\utilde{\epsilon }}, \label{LR}
\end{equation}
$\lambda_{v}$ and $\mu_{v}$ are viscoelastic coefficients. 

The elastic coefficients have the following values (Bauer et al., 2005; Barclay
Satterfield and Benziger, 2009; Silberstein and Boyce, 2010), which are in agreement with the usual ones
\begin{equation}
G\sim 4.5\;10^{7}\;Pa, \qquad \lambda \sim 3\;10^{8}\;Pa, \qquad E\sim
1.3\;10^{8}\;Pa, \qquad \nu \sim 0.435,
\end{equation}
where $E$ is the Young's modulus and $\nu$ the Poisson's ratio. Viscoelastic coefficients 
can be deduced from uniaxial tension tests (Barclay Satterfield and Benziger, 2009;
Silberstein and Boyce, 2010; Silberstein et al., 2011) and relaxation times for a traction, which are close to $15\;s$ 
(Silberstein, 2008; Silberstein and Boyce, 2010;
Silberstein et al., 2011)
\begin{equation}
\lambda_v \sim \; 7 \; 10^8\; Pa \: s,  \qquad \mu_v \sim 10^{8}\;Pa \: s.
\end{equation}
These coefficients depend strongly on the solvent concentration and on the temperature, 
especially if it is close to the glass transition.

\subsection{Nafion$^{\textregistered}$ Physicochemical Properties}
\label{subsec:52}
Vectorial constitutive equations require nine phenomenogical coefficients, which are a 
priori second-rank tensors; they can be replaced by scalars because of the isotropy of 
the medium. The first equation that we obtain is a generalized Fourier's law. We will approximate 
the two other by restricting ourselves to the isothermal case and focusing on a particular 
E.A.P. : Nafion$^{\textregistered}$ 117 $Li^{+}$.

The liquid phase is a dilute solution of strong electrolyte. According to 
Diu et al.
(2007), mass chemical potentials can be written on a first approximation%
\begin{equation}
\begin{tabular}{l}
$\mu _{1}\left( T,p,x\right) \simeq \mu _{1}^{0}\left( T,p\right) +\frac{RT}{M_{1}}%
\ln \left( C\frac{M_{2}}{\rho _{2}^{0}}\right) $, \\ 
$\mu _{2}\left( T,p,x\right) \simeq \mu _{2}^{0}\left( T,p\right) -\frac{RT}{\rho _{2}^{0}}C$, \\ 
$\mu _{3}\left( T,p,x\right) =\mu _{3}^{0}\left( T\right) $, %
\end{tabular}%
\end{equation}%
where $R=8,314\;J\;K^{-1}$ is the gas constant. $\mu _{2}^{0}$ and $\mu _{3}^{0}$ denote 
the chemical potentials of the single solid and solvent, and $\mu _{1}^{0}$ depends on 
the solvent and the solute. The Gibbs-Duhem's relations for the solid and the liquid phases 
enable the calculation of $\vec{grad}\mu_{k}$
\begin{equation}
\begin{tabular}{l}
$\vec{grad}\mu _{1}=-\frac{S_{1}}{\rho _{1}}\vec{grad}T+\frac{v_{1}}{M_{1}}\vec{grad} p
+\frac{RT\rho _{2}^{0}}{M_{2}M_{1}C}\vec{grad}\left( \frac{CM_{2}}{\rho _{2}^{0}}\right) $, \\ 
$\vec{grad}\mu _{2}=-\frac{S_{2}}{\rho _{2}}\vec{grad} T+\frac{v_{2}}{M_{2}}
\vec{grad} p-\frac{RT}{M_{2}}\vec{grad}\left( \frac{CM_{2}}{\rho _{2}^{0}}\right) $, \\ 
$\vec{grad}\mu _{3}=-\frac{S_{3}}{\rho _{3}}\vec{grad}T $,
\end{tabular}%
\end{equation}%
where $v_{k}$ denotes the partial molar volume of the constituent $k$.

The physicochemical properties of the Nafion$^{\textregistered}$ are well documented. Its equivalent weight, 
that is to say, the weight of polymer per mole of ionic sites is $M_{eq} \sim 1.1 
\;kg \; eq^{-1}$ (Gebel, 2000). The solution volume fraction $\phi _{4}$ is close 
to $38\%$ (Cappadonia et al., 1994; Choi et al., 2005). Other parameters are resumed in table \ref{tab:3} 
(Heitner-Wirguin, 1996; Nemat-Nasser and Li, 2000). The anions molar concentration, which is equal 
to the average cations concentration, 
is $C_{moy}=\frac{\rho _{3}^{0} \phi_{3}}{M_{eq}\phi _{4}}\sim 3.1\;10^{3}\;mol\;m^{-3}$. The 
dynamic viscosity of water is $\eta _{2}=10^{-3}\;Pa\;s$.  
We deduce the mass densities of the complete material $\rho \sim 1.7\;10^{3}\;kg\;m^{-3}$. 
We moreover suppose that the temperature is $T=300\;K$. The electric field is 
typically about $10^{4}\;V\;m^{-1}$ (Nemat-Nasser and Li, 2000).

\begin{table}
\caption{Nafion$^{\textregistered}$ 117 $Li^{+}$ parameters}
\label{tab:3}
\begin{tabular}{|ll|l|l|l|}
\hline
&  & Cations & Solvent & Solid \\ \hline
$M_{k}$ & $\left( kg\;mol^{-1}\right) $ & $\mathit{3}\;\mathit{10}^{\mathit{-3}}$ & $%
\mathit{18}\;\mathit{10}^{\mathit{-3}}$ & $\mathit{10}^{\mathit{2}}$ - $%
\mathit{10}^{\mathit{3}}$ \\ \hline
$\rho _{k}^{0}$ & $\left( kg\;m^{-3}\right) $ &  & $\mathit{10}^{\mathit{3}}$
& $\mathit{2.08}\;\mathit{10}^{\mathit{3}}$ \\ \hline
$v_{k}$ & $\left( m^{3}\;mol^{-1}\right) $ & $\frac{M_{1}}{\rho _{4}^{0}}%
\sim \mathit{3}\;\mathit{10}^{\mathit{-6}}$ & $\mathit{18}\;\mathit{10}^{\mathit{-6}}$ & 
\\ \hline
$\rho _{k}$ & $\left( kg\;m^{-3}\right) $ & $\mathit{3.5}$ & $\mathit{380}$ & $\mathit{1.3
\;10}^{3}$ \\ \hline
$Z_{k}$ & $\left( C\;kg^{-1}\right) $ & $\mathit{3.2}\;\mathit{10}^{\mathit{7}}$ & $%
\mathit{0}$ & $\mathit{9\;10}^{\mathit{4}}$ \\ \hline
\end{tabular}%
\end{table}%

Considering this numerical estimations, we can write in a first approximation
\begin{equation}
Z_{1}>>Z_{3}, \qquad \qquad \rho \sim \rho _{2}\sim \rho _{3}>>\rho _{1}, \qquad
\qquad \rho _{1}Z_{1}\sim \rho _{4}Z_{3}.
\end{equation}

\subsection{Nernst-Planck Equation}
\label{subsec:53}

It is commonly accepted that the non-diagonal phenomenological coefficients are small
compared to the diagonal ones. The expression of the mass diffusion flux of the cations 
$J_{1}$ can be identified with a Nernst-Planck equation (Lakshminarayanaiah,
1969)
\begin{equation}
\vec{V_{1}}=-\frac{D}{C}\left[ \vec{grad} C-\frac{Z_{1}M_{1}C}{RT}\vec{E}+\frac{Cv_{1}}{RT}
\left( 1-\frac{M_{1}}{M_{2}}\frac{v_{2}}{v_{1}}\right) \vec{grad} p\right] +\vec{V_{2}}. \label{Nernst}
\end{equation} $D\sim 2\; 10^{-9}m^{2}\;s^{-1}$ 
denotes the mass diffusion coefficient of the cations in the liquid phase (Zawodsinski et al., 1991).
This equation expresses the equilibrium of an ions mole under the action of four forces : 
the Stokes friction force, which is proportional to $\vec{V_{1}}-\vec{V_{2}}$, the pressure 
force, the electric force and the thermodynamic force $-M_{1}\vec{grad} \mu _{1}$.

The order of magnitude of the different terms of this equation can be estimated. According to 
Farinholt and Leo (2004) and Nemat-Nasser (2002), the concentration gradient verifies
\begin{equation}
\left\vert \vec{grad} C\right\vert \precsim 10^{8}\;mol\;m^{-4} \;.
\end{equation}
The pressure gradient can be roughly estimated using the Darcy's law; it is about $10^{9}\;Pa\;m^{-1}$. We thus obtain 
\begin{equation}
\begin{tabular}{ll}
$\frac{M_{1}C}{RT}Z_{1}\left\vert \vec{E}\right\vert $ & $\sim 1.6\;10^{9}\;mol\;m^{-4}$, \\ 
$\frac{Cv_{1}}{RT}\left( 1-\frac{M_{1}}{M_{2}}\frac{v_{2}}{v_{1}}\right)
\left\vert \vec{grad} p\right\vert $ & $\sim1.1\;10^{3}\;mol\;m^{-4}$.
\end{tabular}
\end{equation}
The electric field and the mass diffusion have the leading effects; we thereafter 
neglect the pressure gradient term.

\subsection{Generalized Darcy's Law}
\label{subsec:54}
The expression of the mass diffusion flux of the solution in the solid $J_{4}$ identifies 
with a generalized Darcy's law
\begin{equation}
\vec{V_{4}}-\vec{V_{3}}\simeq -\frac{K}{\eta _{2}\phi_{4}}\left[ \vec{grad} p-\rho _{4}^{0}
\left( Z_{4}-Z_{3}\right) \vec{E} \right] -\frac{R}{M_{1}C \rho_{4}}L^{1}\vec{grad} C.
\end{equation}
where $L^{1}$ is a phenomenological coefficient and $K$ the intrinsic permeability of the 
solid phase, which is on the order of the square of the pore sizes, that is ${10^{-16}} \; m^{2}$.

The orders of magnitude of the different terms are
\begin{equation}
\begin{tabular}{ll}
$\frac{K}{\eta _{2}\phi _{4}}\left\vert \vec{grad} p\right\vert $
& $\sim 2.6\;10^{-4}\;m\;s^{-1}$, \\ 
$\frac{K}{\eta _{2}\phi _{4}} \rho _{4}^{0} \left( %
Z_{4}-Z_{3}\right) \left\vert \vec{E}\right\vert $ & $\sim
0.53\;m\;s^{-1}$, \\ 
$\frac{R}{M_{1}C\rho_{4}}L^{1} \left\vert \vec{grad} C\right\vert $ & $<<5.9\;10^{-7}\;m\;s^{-1}$.%
\end{tabular}%
\end{equation}%
The latter term can therefore be neglected. The first term represents the mass pressure 
force and the second one is the mass electric force; it expresses the motion of the solution 
under the action of the electric field and reflects an electro-osmotic phenomenon.

The distribution of cations becomes very heterogeneous (Farinholt and Leo,
2004; Nemat-Nasser, 2002) : they 
gather near the negative electrode, where $Z_{4}>>Z_{3}$; the expression obtained coincides 
with that of Biot (1955). Near the positive electrode, where the cation concentration is very
low, $Z_{4}<<Z_{3}$; this corresponds to the result obtained by Grimshaw et al. (1990) and Nemat-Nasser and Li (2000). 
In the center of the strip, the mass electric force exerted on the solution is proportional 
to the net charge $\left(Z_{4}-Z_{3}\right) $.

\section{Validation of the Model: Application to a Cantilevered Strip}
\label{sec:6}
In order to validate the model that we have just described, we apply it to the static case of a 
cantilevered E.A.P. strip bending under the effect of a permanent electric field. In addition, 
the strip might undergo the action of a shear force exerted on the free end in order 
to prevent its displacement (blocking force).

\subsection{Static Equations}
\label{subsec:61}

A continuous constant voltage $\varphi_{0}$ is applied between the two faces of the strip. As a 
consequence, the partial derivatives with respect to time and the velocities are zero. 
We focuse on Nafion$^{\textregistered}$ $Li^{+}$ strip of length $L=2~cm$, of thickness $2e=200~\mu m$
and of width $2l=5~mm$ subject to a potential difference $\varphi_{0} = 1~V$. We postulate that 
the volume fraction $\phi_{4}$ is constant; this hypothesis will be checked a posteriori. As a 
consequence, the dielectric permittivity of the whole material is a constant too. We assume that 
it is equal to that measured by Deng and Mauritz (1992) for a material very close to the Nafion$^{\textregistered}$ : 
${\varepsilon \sim 10^{-6}~F~m^{-1}}$. Considering the strip dimensions, it is a two-dimensional 
problem. A coordinate system $Oxyz$ is chosen such that the axis $Oz$ is parallel to the 
imposed electric field, the axis $Ox$ is along the length of the strip and $Oy$ along its width. 
We venture the hypothesis that the local electric field axial coordinate $E_{x}$ is negligible 
compare to the normal one $E_{z}$. On a first approximation, $C$, $E_{z}$, $p$, the local 
electric potential $\varphi$ and the electric charge density $\rho Z$ only depend on the $z$ 
coordinate. Neglecting the pressure term in (\ref{Nernst}), we obtain
\begin{equation}
\begin{tabular}{clcclc}
$E_{z}$&$= - \frac {d \varphi} {dz} $, &$ \qquad \qquad \qquad $ & 
$\varepsilon \frac {d E_{z}} {dz} $&$= \rho Z$, \\
$\frac {dp} {dz} $&$= \left(C F - \rho _{2}^{0} Z_{3}\right) E_{z} $, &$ \qquad \qquad \qquad $ &
$\frac {dC} {dz} $&$= \frac{FC}{RT} E_{z} $, \\
$\rho Z $&$= \phi_{4} F \left(C - C_{moy} \right)$. & $ $ & $ $
\end{tabular}
\end{equation}
$F=96487\;C\;mol^{-1}$ is the Faraday's constant. The boundary conditions write
\begin{equation}
\begin{tabular}{ccc}
$\underset{z \rightarrow -e}{lim} \varphi =\varphi_{0}, \qquad \qquad $ & $%
\underset{z \rightarrow e}{lim} \varphi = 0, \qquad \qquad $ & $
\int_{-e}^{e} \rho Z \, \mathrm{d}z =0 $.
\end{tabular} \label{CL}
\end{equation}
The latter condition expresses the electroneutrality condition and is equivalent to $E_{z}\left(
 e \right) = E_{z}\left( -e \right)$. We introduce the following dimensionless variables:
\begin{equation}
\begin{tabular}{clclclc}
$\overline{E} $&$= \frac {E_{z} e} {\varphi_{0}}, \qquad \qquad $ & $ \overline{C} $&$= \frac {C} 
{C_{moy}}, \qquad \qquad $ & $ \overline{\varphi} $&$= \frac {\varphi} {\varphi_{0}},$ \\
$\overline{\rho Z} $&$= \frac {\rho Z} {\phi_{4} F C_{moy}}, \qquad \qquad $ & $ 
\overline{p} $&$= \frac {p} {F \varphi_{0} C_{moy}}, \qquad \qquad $ & $ \overline{z} $&$= \frac {z} {e}$, \\
\end{tabular}
\end{equation}
which leads to
\begin{equation}
\begin{tabular}{clclc}
$\overline{E}$&$= - \frac {d \overline{\varphi}} {d \overline{z}} $, & $ \qquad \qquad \qquad $ & 
$\frac {d \overline{E}} {d \overline{z}}$&$ = \frac {A_{1}} {A_{2}}\overline{\rho Z} $, \\
$\frac {d \overline{p}} {d\overline{z}}$&$ = \left(\overline{C} + A_{3} \right) \overline{E} 
$, & $ \qquad \qquad \qquad $ &
$\frac {d \overline{C}} {d \overline{z}}$&$ = A_{2} \overline{C} \overline{E} $, \\
$\overline{\rho Z} $&$= \overline{C} - 1. $ & $ $ & $ $
\end{tabular} \label{Eqp}
\end{equation}
$A_{1}$, $A_{2}$ et $A_{3}$ are dimensionless constants
\begin{equation}
\begin{tabular}{ccc}
$A_{1}=\frac {\phi_{4} e^{2} F^{2} C_{moy}} {\varepsilon R T} \sim 4,37~10^{7}, \; $ &
$A_{2}=\frac { F \varphi_{0}} {R T} \sim 38,7, \; $ &
$A_{3}= -\frac {\rho_{2}^{0} Z_{3} C_{moy}} {F C_{moy}} \sim 0,303 $.
\end{tabular}
\end{equation}
Boundary conditions are
\begin{equation}
\begin{tabular}{ccc}
$\underset{\overline{z} \rightarrow -1}{lim} \overline{\varphi} = 1, \qquad \qquad $ & $%
\underset{\overline{z} \rightarrow 1}{lim} \overline{\varphi} = 0, \qquad \qquad $ & $
\overline{E}\left( 1 \right) = \overline{E}\left( -1 \right) $.
\end{tabular} \label{CLa}
\end{equation}
We deduce
\begin{equation}
\frac {d} {d\overline{z}} \left( \frac {d \overline{C}} {\overline{C} d \overline{z}} \right) 
= A_{1} \left( \overline{C} - 1  \right), \label{E2bis}
\end{equation}
and
\begin{equation}
\begin{tabular}{ccc}
$\overline{C} \simeq A_{2} exp \left(-A_{2} \overline{\varphi} \right) $ & $, \qquad \qquad \qquad $ &  
$ \overline{p} = \frac {\overline{C}} {A_{2}} - A_{3} \overline{\varphi} + Cte $.
\end{tabular} \label{E5E6}
\end{equation}

The hydrated polymer can be assimilated to a conductive material. As a consequence, the electric 
field is zero throughout the strip except near the boundaries. We deduce the values of the different 
quantities on the sides and at the center of the strip (table \ref{tab:4}).
\begin{table}
\caption{Boundary values of the physical quantities}
\label{tab:4}
\begin{tabular}{|c|c|c|c|}
\hline
& $-1$ & $Center$ & $1$ \\ \hline
$\overline{C}$ & $A_{2}e^{-A_{2}}\simeq 0$ & $1$ & $A_{2}$ \\ \hline
$\overline{\varphi }$ & $1$ & $\frac{\ln A_{2}}{A_{2}}$ & $0$ \\ \hline
$\overline{E}$ & $\sqrt{\frac{2A_{1}}{A_{2}}\left[ 1-\frac{1}{A_{2}}\left( 1+\ln A_{2}\right)
\right] }$ & $0$ & 
$\sqrt{\frac{2A_{1}}{A_{2}}\left[ 1-\frac{1}{A_{2}}\left( 1+\ln A_{2}\right)\right] }$ \\ \hline
$\overline{p}$ & $-A_{3}$ & $\frac{1}{A_{2}}\left( 1-A_{3}\ln
A_{2}\right) $ & $1$ \\ \hline
\end{tabular}%
\end{table}
The equation (\ref{E2bis}) can be solved using Matlab. We deduce $\overline{\rho Z}$, 
$\overline{\varphi}$, $\overline{p}$ and $\overline{E}$ by (\ref{Eqp}) and (\ref{E5E6}). An evaluation of the 
pressure term of the equation (\ref{Nernst}) shows that it does not exceed $2\%$ of the second 
term of the equation with the nominal conditions chosen.

\subsection{Beam Model on Large Displacements}
\label{subsec:62}

\begin{wrapfigure}{l}{0.5\textwidth}
\includegraphics [width=\linewidth]{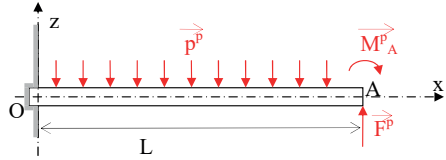}
\caption{Forces exerted on the beam}
\label{fig:2}
\end{wrapfigure}

Given the high deflection values, we used a large displacement beam model to determine forces, 
stress and strain. We consider a straight beam clamped at the end $O$; the other end $A$ is either 
free or subject to a shear force bringing the deflection to a zero value (blocking force 
$\overrightarrow{F^{p}}$). The polymer is subject to a distributed electric force 
$\overrightarrow{p^{p}}$ independent of the $x$ coordinate and orthogonal to the strip. Moreover, 
the swelling of the strip on the side of the negative electrode generates, through the pressure 
$p=\frac {\sigma_{xx}} {3}$, a bending moment $\overrightarrow{M_{A}^{p}}$ at the free end of the beam (see figure 2). 
Considering the electroneutrality condition, the distributed force and the bending moment verify
\begin{equation}
p^{p} = \int_{-l}^{l} \int_{-e}^{e} \rho Z E_{z}~dz~dy = 2l \int_{-e}^{e} \rho Z E_{z}~dz
= 2l \varepsilon \left[ \frac{E_{z}^{2}}{2} \right] _{-e}^{e} = 0,
\end{equation} 
\begin{equation}
M_{A}^{p}= \int_{-l}^{l} \int_{-e}^{e}\sigma_{xx}~z~dz~dy = A_{4}\int_{-1}^{1}\overline{p}\;
\overline{z}\;d\overline{z},
\end{equation}
with $A_{4}=6le^{2}F\varphi _{0}C_{moy}\sim 0,045\;N\;m$.  \\

\begin{wrapfigure}{r}{0.36\textwidth}
\begin{center}
\vspace{-1.5 cm}
\includegraphics [width=0.38\textwidth]{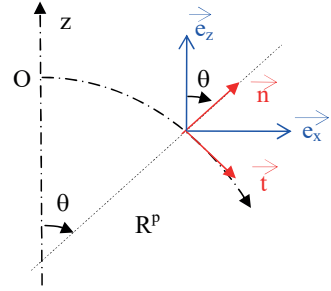}
\caption{Beam on large displacements : geometric coordinates}
\label{fig:3}
\end{center}
\end{wrapfigure}

We assume that the Bernoulli and Barr\'{e} Saint Venant hypotheses are verified. Let $s$ and 
$\overline{s}$ be the curvilinear abscissas along the beam respectively at the 
rest and deformed configurations, 
$\overrightarrow{t}$ and $\overrightarrow{n}$ the vectors tangent and normal to the beam, 
$\theta$ the angle of rotation of a cross-section and $R^{p}$ the radius of curvature (geometric 
parameters are given in figure 3). It will be assumed that the elongation 
$\Lambda = \frac{d \overline{s}}{ds}$ is equal to $1$. The bending moment in the current section is
\vspace{-0.35 cm}
\begin{equation}
M^{p} = F^{p} \left( x-L \right) + M_{A}^{p}.
\end{equation}
The strain tensor is given by
\begin{equation}
\utilde{\epsilon} = \frac{1}{2} \left[ \left( \utilde{Grad} \vec{u} + \utilde{1} \right) ^{T} 
\left( \utilde{Grad} \vec{u} + \utilde{1} \right) - \utilde{1} \right].
\end{equation}
In the reference frame linked to the undeformed beam, it follows
\begin{equation}
\epsilon_{xx} = - \frac{\overline{n}}{R^{p}} \left( 1 - \frac{\overline{n}}{2R^{p}} \right) \simeq - 
\frac{\overline{n}}{R^{p}},
\end{equation}
where $\overline{n}$ designates the coordinate in the $\vec{n}$ direction. Indeed, the beam 
being thin, $\left|\overline{n}\right|<<R^{p}$.
In the case of pure bending, the strain is given by $\epsilon_{xx}=\frac{{M}^{p}} {EI^{p}}\overline{n}$
where $ I^{p}=\frac{4le^{3}}{3}$ is the area moment of inertia, with respect to the $Oy$ axis. 
The shear force has a negligible effect on the deflection. We deduce the radius of curvature 
and the angle of rotation
\begin{equation}
\begin{tabular}{cc}
$\frac{1}{R^{p}} = \frac{d \theta}{d \overline{s}} = \frac{F^{p}}{E I^{p}}  \left( L-\overline{s} 
\right) - \frac{M^{p}_{A}}{E I^{p}}, \qquad \qquad $
$\qquad \theta = \frac{F^{p}}{2E I^{p}} \overline{s} \left( 2L-\overline{s} \right) - 
\frac{M^{p}_{A}}{E I^{p}} \overline{s} $,
\end{tabular}
\end{equation}
by choosing the point $O$ as the origin of the curvilinear abscissas. The deflection $w$ 
is obtained by integrating the relation $\frac {dz} {d\overline{s}}=sin\theta$.
In the case of a cantilever beam ($F^{p}=0$), the radius of curvature is constant; 
the beam is circle shaped and at the free end
\begin{equation}
\begin{tabular} {l}
$\theta = - \frac{M_{A}^{p}}{E I^{p}} L = -\frac{2A_{5}}{L}\int_{-1}^{1}\overline{p}\;\overline{z}
\;d \overline{z} $, \\
$w = \frac{E I^{p}}{M_{A}^{p}} \left[ cos \left( \frac{M_{A}^{p}}{E I^{p}} L \right) -1 \right]
=\frac{L^{2}}{2A_{5}\int_{-1}^{1}\overline{p}\;\overline{z}\;d \overline{z}}\left[ \cos \left( 
\frac{2A_{5}}{L}\int_{-1}^{1}\overline{p}\;\overline{z}\;d\overline{z}\right) -1\right] $,
\end{tabular}
\end{equation}
with $A_{5}=\frac{9}{4}\frac{L^{2}F\varphi _{0}C_{moy}}{eE}\sim 20,59\;m$.
The blocking force is the end loading such that the deflection is zero

\begin{equation}
w \left( \overline{s}=L \right) = - \int_{0}^{L} \sin \left[ \frac{F^{p}}{2EI^{p}} 
\overline{s} \left( \overline{s}-2L \right) + \frac{M_{A}^{p}}{EI^{p}} 
\overline{s} \right] \; d \overline{s} = 0.
\end{equation}

Using the Fresnel functions to compute this integral, we obtain for the blocking 
force the same result as in small displacements
\begin{equation}
F^{p} = \frac{3 M_{A}^{p}}{2L}=\frac{3A_{4}}{2L}\int_{-1}^{1} \overline{p}\;\overline{z}\;d\overline{z}.
\end{equation}

\subsection{Simulations Results}
\label{subsec:63}
The values obtained for the deflection and the blocking force are in good agreement with the 
experimental values reported in the literature (Nemat-Nasser, 2002;
Newbury, 2002; Newbury and Leo, 2002, 2003).

\begin{figure*} [t]
 \includegraphics[width=0.9\textwidth]{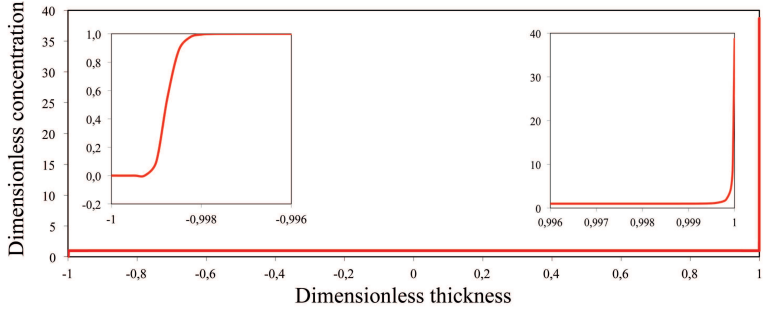}
\caption{Variation of cation concentration in the thickness of the strip; the distributions 
at the vicinity of the lower and upper faces (electrodes) are detailled in insets.}
\label{fig:4}
\end{figure*}
The variation of cation concentration in the thickness of the strip is shown in figure \ref{fig:4}. 
This quantity is constant throughout the central part of the strip but varies very strongly in the 
vicinity of the electrodes, especially near the negative electrode on which the cations accumulate. 
On the contrary, it is noticed near the positive electrode an aerea of the order of $0,1~\mu m$ 
without cations. This result is in good agreement with that of Nemat-Nasser (2002).\\

\begin{wrapfigure}{r}{0.5\textwidth}
\vspace{-1cm}
\includegraphics [width=\linewidth]{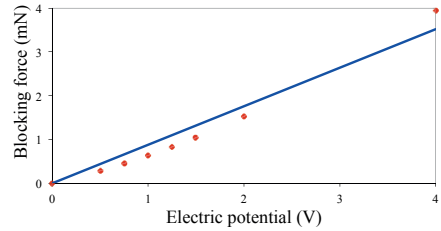}
\caption{Influence of the potential difference on the blocking force}
\label{fig:5}
\end{wrapfigure}

$\int_{-1}^{1}\overline{p}\;\overline{z}\;d \overline{z}$ depends only on the imposed potential 
$\varphi_{0}$, the thickness $e$ and the cation chosen. The deflection $w$ is therefore independent 
of the width and approximately proportional to length square, which corresponds to Shahinpoor's 
measurements (Shahinpoor,
1999). The blocking force is proportional to $l$ and inversely proportional 
to $L$, which is in good agreement with Newbury and Leo (2003).

The deflection varies 
linearly with the imposed potential difference $\varphi_{0}$, a result in 
agreement with the experiments of Mojarrad and Shahinpoor (1997) and Shahinpoor
et al. (1998). The blocking force follows the same tendency (figure \ref{fig:5}).

\section{Conclusion}
\label{sec:7}
We have modeled the behavior of an ionic electroactive polymer saturated with water and 
subject to an orthogonal electric field. The presence of water causes a complete dissociation 
of the polymer and the release of small cations. We have depicted this medium as the superposition 
of three systems with different velocities fields : the cations, the solvent and the solid 
assimilated to a deformable porous medium. We have used a continuous medium approach and a coarse
grain model. We have established the microscale balance equations of mass, linear momentum, 
energies, entropy and the Maxwell's equations as well as thermodynamic relations for each 
phase (solid and liquid). We have derived the macroscale equations relative to the whole material 
using an average technique and the material derivative concept. Thermodynamics of linear 
irreversible processes have provided the entropy production and the constitutive equations of 
the complete material. We have obtained a rheological law of the Kelvin-Voigt type, generalised 
Fourier's and Darcy's laws and a Nernst-Planck equation.

We have applied this model to a cantilevered Nafion$^{\textregistered}$ strip subject to a continuous voltage 
between its two faces, which is a static case. The other end may be either free or subject 
to a shear force preventing its displacement (blocking force). We have used a beam model in large 
displacement. We have drawn the profile of cations concentration and evaluated the deflection 
and the blocking force. We have observed that the concentration varies very 
strongly in the vicinity of the 
electrodes, which is characteristic of a conductive material behavior. Our results are in 
good agreement with the experimental data published in the literature.

To improve this model, we intend to take into account the variations of the permittivity with the 
cation concentration. Another way of improvement is to replace the rheological law with a 
Zener model that is better suited to viscoelastic polymers.

\section{Notations}
\label{sec:8}
$k=1,2,3,4,i$ subscripts respectively represent cations, solvent, solid,
solution (water and cations) and interface; quantities without subscript
refer to the whole material. Superscript $^{0}$ denotes a local quantity;
the lack of superscript indicates average quantity at the macroscopic\
scale. Microscale volume quantities are relative to the volume of the phase,
average quantities to the volume of the whole material. Superscript $^{s}$ 
indicates the deviatoric part of a second-rank tensor, and $^{T}$ its transpose.

\begin{description}
\item $C$, $C_{moy}$ : cations molar concentrations (relative to the liquid phase);

\item $D$ : mass diffusion coefficient of the cations in the liquid phase;

\item $\vec{D}$, $\vec{D_k^0}$  : electric displacement field;

\item $e$ : half-thickness of the strip;

\item $E$, $G$, $\lambda$, $\nu$ : Young's and shear modulus, first Lam\'{e} constant, Poisson's ratio;

\item $\vec{E}$, $\vec{E_k^0}$ : electric field;

\item $E$, $E_{p}$ ($E_{pk}^{0}$), $E_{c}$, $U$ ($U_{k}$, $U_{k}^{0}$) : total, potential, kinetic and internal energy densities;

\item $F=96487\;C\;mol^{-1}$ : Faraday's constant ;

\item $\vec{F^{p}}$ : blocking force;

\item $\vec{i}$ ($\vec{i^{\prime }}$) : diffusion current;

\item $\vec{I}$ ($\vec{I_{k}}$, $\vec{I_{k}^{0}}$) : current density vectors;

\item $I^{p}$ : area moment of inertia;

\item $\vec{J_{k}}$ : mass diffusion flux;

\item $K$ : intrinsic permeability of the solid phase;

\item $l$ : half-width of the strip;

\item $L$ : length of the strip;

\item $M_{k}$ : molar mass of component $k$;

\item $M_{eq}$ : equivalent weight (weight of polymer per mole of sulfonate
groups);

\item $\vec{M^{p}}$ ($\vec{M_{A}^{p}}$) : bending moment;

\item $\vec{n_{k}}$ : outward-pointing unit normal of phase $k$;

\item $p$ ($p_{k}$, $p_{k}^{0}$) : pressure;

\item $\vec{Q}$ ($\vec{Q^{\prime }}$) : heat flux;

\item $R=8,314\;J\;K^{-1}$ : gaz constant;

\item $R^{p}$ : radius of curvature of the beam;

\item $s$ : rate of entropy production;

\item $S$ ($S_{k}^{0}$) : entropy density;

\item $T$ ($T_{k}$, $T_{k}^{0}$) : absolute temperature;

\item $\vec{u}$ : displacement vector;

\item $v_{k}$ : partial molar volume of component $k$ (relative to the
liquid phase);

\item $\vec{V}$ ($\vec{V_{k}}$, $\vec{V_{k}^{0}}$) : velocity;

\item $w$ : deflection of the beam;

\item $Z$ ($Z_{k}$) : total electric charge per unit of mass;

\item $\varepsilon$, $\varepsilon_k^0$ : permittivity;

\item $\utilde{\epsilon}$ ($\utilde{\epsilon_{k}}$, $\utilde{\epsilon_{k}^{0}}$) : strain tensor;

\item $\eta_{2}$ : dynamic viscosity of water;

\item $\theta$ : angle of rotation of a beam cross section;

\item $\lambda_{v}$, $\mu_{v}$ : viscoelastic coefficients;

\item $\mu_{k}$ ($\mu_{k}^{0}$) : mass chemical potential;

\item $\rho $ ($\rho _{k}$, $\rho _{k}^{0}$) : mass density;

\item $\utilde{\sigma}$ ($\utilde{\sigma_{k}}$, $\utilde{\sigma_{k}^{0}}$), $\utilde{\sigma^{v}}$, 
$\utilde{\sigma^{e}}$ ($\utilde{\sigma _{k}^{e}},\utilde{\sigma_{k}^{0e}}$) : stress tensor, dynamic 
stress tensor, equilibrium stress tensor;

\item $\pmb{\Sigma}$ : entropy flux vector;

\item $\phi_{k}$ : volume fraction of phase $k$;

\item $\varphi$ ($\varphi_{0}$): electric potential;

\item $\chi_{k}$ : function of presence of phase $k$ ;
\end{description}

\section*{References}

\begin{enumerate} 

\item[] \hskip -0.5cm A. Ask, A. Menzel, and M. Ristinmaa. Electrostriction in electro-viscoelastic polymers. \textit{Mechanics of Materials,} 50:9–21, 2012.

\item[] \hskip -0.5cm Y. Bar-Cohen. Electroactive polymers (eap) actuators artificial muscles : reality, potential and challenges.\textit{ SPIE Press,} 2001.

\item[] \hskip -0.5cm M. Barclay Satterfield and J. B. Benziger. Viscoelastic properties of nafion at elevated temperature and humidity. \textit{J. Polym. Sci. Pol. Phys.,} 47(1):11–24, 2009.

\item[] \hskip -0.5cm F. Bauer, S. Denneler, and M. Willert-Porada. Influence of temperature and humidity on the mechanical properties of nafion$^{\textregistered}$ 117 polymer electrolyte membrane. \textit{Journal of Polymer Science part B : Polymer Physics,} 43(7):786–795, 2005.

\item[] \hskip -0.5cm M. A. Biot. Theory of elesticity and consolidation for a porous anisotropic solid. \textit{Journal of Applied Physics,} 26(2):182–185, 1955.

\item[] \hskip -0.5cm Variational lagrangian-thermodynamics of nonisothermal finite strain mechanics of porous solids and thermomolecular diffusion. \textit{International Journal of Solids and Structures,} 13:579–597, 1977.

\item[] \hskip -0.5cm J. Brafau-Penella, M. Puig-Vidal, P. Giannone, S. Graziani, and S. Strazzeri. Characterization of the harvesting capabilities of an ionic polymer metal composite device. \textit{Smart material and Structures,} 17(1):015009.1–015009.15, 2008.

\item[] \hskip -0.5cm M. Cappadonia, J. Erning, and U. Stimming. Proton conduction of nafion$^{\textregistered}$ 117 membrane between 140 k and room temperature. \textit{Journal of Electroanalytical Chemistry,} 376(1):189–193, 1994.

\item[] \hskip -0.5cm J. Chabé. \textit{Etude des interactions moléculaires polymère-eau lors de l’hydratation de la membrane Nafion, électrolyte de référence de la pile combustible.} Phd thesis, Université Joseph Fourier Grenoble I, 2008. $http ://tel:archives-ouvertes.fr/docs/00/28/59/99/PDF/THESE_JCS.pdf$.

\item[] \hskip -0.5cm P. Choi, N.H. Jalani, and R. Datta. Thermodynamics and proton transport in nafion : Ii. proton diffusion mechanisms and conductivity. \textit{Journal of the Electrochemical Society,} 152(3):E123–E130, 2005.

\item[] \hskip -0.5cm F. Collette. \textit{Vieillissement hygrothermique du Nafion.} Phd thesis, Ecole Nationale Supérieure d’Arts et Métiers, 2008. \textit{$http://tel.archives-ouvertes.fr/docs/00/$ $35/48/47/PDF/These\_Floraine$ $\_COLLETTE\_27112008.pdf$}

\item[] \hskip -0.5cm O. Coussy. Dielectric relaxation studies of water-containing short side chain perfluorosulfonic acid membranes. \textit{Eur. J. Mech.} A, 8(1):1–14, 1989.

\item[] \hskip -0.5cm O. Coussy. \textit{Mechanics of porous continua.} Wiley, Chichester, 1995.

\item[] \hskip -0.5cm P. de Gennes, K. Okumuro, M. Shahinpoor, and Ki K.J. Mechanoelectric effects in ionic gels. \textit{Europhysics Letters}, 50(4):513–518, 2000.

\item[] \hskip -0.5cm S. R. de Groot and P. Mazur. \textit{Non-equilibrium thermodynamics.} North-Holland publishing company, Amsterdam, 1962.

\item[] \hskip -0.5cm Z.D. Deng and K.A. Mauritz. Dielectric relaxation studies of water-containing short side chain perfluorosulfonic acid membranes. \textit{Macromolecules,} 25(10):2739–-2745, 1992.

\item[] \hskip -0.5cm B. Diu, C. Guthmann, D. Lederer, and B. Roulet. \textit{Thermodynamique.} Hermann, Paris, 2007.

\item[] \hskip -0.5cm D.A. Drew. Mathematical modeling of two-phase flow. \textit{Annual Review of Fluid Mechanics,} 15(1):261–291, 1983.

\item[] \hskip -0.5cm D.A. Drew and S.L. Passman. \textit{Theory of Multicomponents Fluids.} Springer, New-York, 1998.

\item[] \hskip -0.5cm K. Farinholt and D.J. Leo. Modeling of electromechanical charge sensing in ionic polymer transducers. \textit{Mechanics of Materials,} 36(5):421–433, 2004.

\item[] \hskip -0.5cm G. Gebel. Structural evolution of water swollen perfluorosulfonated ionomers from dry membrane to solution. \textit{Polymer,} 41(15):5829–5838, 2000.

\item[] \hskip -0.5cm T.D. Gierke, G.E. Munn, and F.C. Wilson. The morphology in nafion perfluorinated membrane products, as determined by wide- and small-angle x-ray studies. \textit{Journal of Polymer Science : Polymer Physics Edition,} 19(11):1687–-1704, 1981.

\item[] \hskip -0.5cm P.E. Grimshaw, J.H. Nussbaum, A.J. Grodzinsky, and M.L. Yarmush. Kinetics of electrically and chemically induced swelling in polyelectrolyte gels. \textit{The Journal of Chemical Physics,} 93(6):4462–4472, 1990.

\item[] \hskip -0.5cm C. Heitner-Wirguin. Recent advances in perfluorinated ionomer membranes : structure, properties and applications. \textit{Journal of Membrane Science,} 120(1):1–33, 1996.

\item[] \hskip -0.5cm M. Ishii and T. Hibiki. \textit{Thermo-fluid dynamics of two-phase flow.} Springer, New-York, 2006.

\item[] \hskip -0.5cm J.D. Jackson. \textit{Classical Electrodynamics.} Wiley \& sons, New-York, 1975.

\item[] \hskip -0.5cm N. Lakshminarayanaiah. \textit{Transport Phenomena in Membranes.} Academic Press, New-York, 1969.

\item[] \hskip -0.5cm D. Lhuillier. A mean-field description of two-phase flows with phase changes. \textit{International Journal of Multiphase Flow,} 29(3):511–525, 2003.

\item[] \hskip -0.5cm G.A. Maugin. \textit{Continuum mechanics of electromagnetic solids.} North-Holland, Amsterdam, 1988.

\item[] \hskip -0.5cm M. Mojarrad and M. Shahinpoor. Ion-exchange-metal composite sensor films. \textit{Proc. SPIE,} 3042:52–60, 1997.

\item[] \hskip -0.5cm S. Nemat-Nasser. Micromechanics of actuation of ionic polymer-metal composites. \textit{Journal of Appied Physics,} 92(5):2899–2915, 2002.

\item[] \hskip -0.5cm S. Nemat-Nasser and J. Li. Electromechanical response of ionic polymer-metal composites. \textit{Journal of Appied Physics,} 87(7):3321–3331, 2000.

\item[] \hskip -0.5cm K.M. Newbury. \textit{Characterization, modeling and control of ionic-polymer transducers.} Phd thesis, Faculty of the Virginia Polytechnic Institute and State University, Blacksburg, Virginia, 2002.

\item[] \hskip -0.5cm K.M. Newbury and D.J. Leo. Mechanical work and electromechanical coupling in ionic polymer bender actuators. \textit{In ASME Adative Structures and Materials Symposium,} 2001.

\item[] \hskip -0.5cm K.M. Newbury and D.J. Leo. Electromechanical modeling and characterization of ionic polymer benders. \textit{Journal of Intelligent Material Systems and Structures,} 13(1):51–60, 2002.

\item[] \hskip -0.5cm K.M. Newbury and D.J. Leo. Linear electromechanical model of ionic polymer transducers - part ii : Experimental validation. \textit{Journal of Intelligent Material Systems and Structures,} 14(6):343–357, 2003.

\item[] \hskip -0.5cm R.I. Nigmatulin. Spatial averaging in the mechanics of heterogeneous and dispersed systems. \textit{Int. J. Multiphase Flow,} 5:353–385, 1979.

\item[] \hskip -0.5cm R.I. Nigmatulin. \textit{Dynamics of multiphase media, volume 1 and 2.} Hemisphere, New-York, 1990.

\item[] \hskip -0.5cm M. Shahinpoor. Continuum electromechanics of ionic-polymer gels as artificial muscles for robotic applications. \textit{Smart Materials and Structures,} 3:367–372, 1994.

\item[] \hskip -0.5cm M. Shahinpoor. Electromechanics of ionoelastic beams as electrically controllable artificial muscles. \textit{Proc. of SPIE,} 3669:109–121, 1999.

\item[] \hskip -0.5cm M. Shahinpoor, Y. Bar-Cohen, J.O. Simpson, and J. Smith. Ionic polymer-metal composites (ipmcs) as biomimetic sensors, actuators and artificial muscles - a review. \textit{Smart Materials and Structures,} 7(6):R15–R30, 1998.

\item[] \hskip -0.5cm M. N. Silberstein and M. C. Boyce. Constitutive modeling of the rate, temperature and hydration dependent deformation response of nafion to monotonic and cyclic loading. \textit{J. Power Sources,} 195(17):5692–5706, 2010.

\item[] \hskip -0.5cm M. N. Silberstein, P. V. Pillai, and M. C. Boyce. Biaxial elastic-viscoplastic behaviour of nafion membranes. \textit{Polymer,} 52(2):529–539, 2011.

\item[] \hskip -0.5cm M.N. Silberstein. \textit{Mechanics of Proton Exchange Membranes : Time, Temperature and Hydration Dependence of the Stress-Strain Behavior of Persulfonated Polytetrafluoroethylene.} Phd thesis, Massachusetts Institut of Technology, Cambridge, MA, 2008.

\item[] \hskip -0.5cm M. Tixier and J. Pouget. Conservation laws of an electro-active polymer. \textit{Continuum Mechanics and Thermodynamics,} 26(4):465–481, 2014.

\item[] \hskip -0.5cm M. Tixier and J. Pouget. Constitutive equations for an electroactive polymer. \textit{Continuum Mechanics and Thermodynamics,} 28(4):1071–1091, 2016.

\item[] \hskip -0.5cm S. Todokoru, S. Yamagami, T. Takamori, and K. Oguro. Modeling of nafion-pt composite actuators (icpf) by ionic motion. \textit{Proc. of SPIE,} 3987, 2000.

\item[] \hskip -0.5cm D. Vokoun, He Qingsong, L. Heller, Y. Min, and D.D. Zhen. Modeling of ipmc cantilever’s displacements ans blocking forces. \textit{Journal of Bionic Engineering,} 12(1):142–151, 2015.

\item[] \hskip -0.5cm W.J. Yoon, P.G. Reinhall, and E.J. Seidel. Analysis of electroactive polymer bending : A component in a low cost ultrathin scanning endoscope. \textit{Sensors and Actuators A,} 133(2):506–517, 2007.

\item[] \hskip -0.5cm T.A. Zawodsinski, M. Neeman, L.O. Sillerud, and S. Gottesfeld. Determination of water diffusion coefficients in perfluorosulfonate ionomeric membranes. \textit{The Journal of Physical Chemistry,} 95(15):6040–6044, 1991.

\end{enumerate}


\end{document}